# MULTI-MICROPHONE COMPLEX SPECTRAL MAPPING FOR SPEECH DEREVERBERATION

*Zhong-Qiu Wang♪ and DeLiang Wang♪,♫*

♪Department of Computer Science and Engineering, The Ohio State University, USA
♫Center for Cognitive and Brain Sciences, The Ohio State University, USA
{wangzhon, dwang}@cse.ohio-state.edu

## ABSTRACT

This study proposes a multi-microphone complex spectral mapping approach for speech dereverberation on a fixed array geometry. In the proposed approach, a deep neural network (DNN) is trained to predict the real and imaginary (RI) components of direct sound from the stacked reverberant (and noisy) RI components of multiple microphones. We also investigate the integration of multi-microphone complex spectral mapping with beamforming and post-filtering. Experimental results on multi-channel speech dereverberation demonstrate the effectiveness of the proposed approach.

## 1. INTRODUCTION

Microphone array processing is pervasive in modern hands-free speech communication systems such as smart speakers and phones. With multiple microphones, spatial information can be leveraged in addition to spectral cues to improve speech enhancement and audio source separation. Conventionally, multi-microphone beamforming followed by monaural post-filtering is the most popular approach for multi-channel speech enhancement [1]–[4]. This approach requires an accurate estimate of target direction, and power spectral density and covariance matrices of speech and noise. The estimation is traditionally performed by using for example non-negative matrix factorization or maximum likelihood estimation under a hypothesized probabilistic distribution. Recently, riding on the development of DNN, time-frequency (T-F) masking and mapping based approaches have been established as the main approaches for single-channel speech enhancement and speaker separation [2], [5]–[7]. These studies suggest that deep learning can substantially improve magnitude estimation. In addition, such mask or magnitude estimation provides a powerful means for acoustic beamforming [8], [9] and sound source localization [10]. As a mask value at a T-F unit is close to one, the phase at that unit is little contaminated. Such T-F units can therefore be utilized to robustly compute the covariance matrix of each source for beamforming and localization.

In the line of research on masking-based beamforming, earlier efforts [8], [10]–[14] only use DNN on spectral features to compute a mask for each microphone, and the estimated masks at different microphones are then pooled together to identify T-F units dominated by the same source across all the microphones for covariance matrix computation. Subsequent studies incorporate spatial features such as inter-channel phase differences (IPD) [15], [16], cosine and sine IPD, target direction compensated IPD [17], beamforming results [18], [19], and stacked phases and magnitudes [20], [21] as a way of leveraging spatial information to further improve mask estimation for beamforming. However, these studies aim at improving mask or magnitude estimation, and do not address phase estimation.

In addition, many studies assume a relatively blind setup, where the trained models are designed to be directly applicable to arrays with any number of microphones arranged in an unknown geometry. Although this flexibility is desirable, in applications such as

This research was supported in part by an NIDCD grant (R01 DC012048), an NSF grant (ECCS-1808932), and the Ohio Supercomputer Center.

Amazon Echo and Google Home, the device only has a fixed microphone array with a known number of microphones and geometry. How to leverage this fixed geometry for robust speech processing is therefore an interesting research problem to investigate.

As an initial step towards multi-channel speech enhancement, this study proposes a multi-microphone complex spectral mapping approach for speech dereverberation based on a fixed array geometry, where the real and imaginary (RI) components of multiple microphones are concatenated as input features for a DNN to predict the RI components of the direct-path signal(s) captured at a reference microphone or at all the microphones. The initially estimated target speech can be utilized to compute a beamformer, and the RI components of the beamforming results can be further combined with the RI components of all the microphone signals for post-filtering.

Why should this approach work? We believe that, for a fixed-geometry array, the neural network could learn to enhance the speech arriving from a specific direction by exploiting the spatial information contained in multiple microphones. This approach is in a way similar to recent studies of classification-based sound source localization for arrays with fixed geometry, where a DNN is trained to learn a one-to-one mapping from the inter-channel phase patterns of multiple microphones to the target direction [22]–[25]. Based on deep learning, the proposed approach has the potential to model the non-linear spatial information contained in multi-microphone inputs, while conventional beamforming is only linear and typically utilizes second-order statistics [1] within each frequency.

Although there are time-domain approaches that use multi-microphone modeling for speech enhancement and source separation [26]–[28], their effectiveness in environments with moderate to strong reverberation is not yet established [29]. In addition, our study tightly integrates multi-microphone complex spectral mapping with beamforming and post-filtering.

The rest of this paper presents the physical model and proposed algorithms in Sections 2 and 3, experimental setup and evaluation results in Sections 4 and 5, and conclusions in Section 6.

## 2. PHYSICAL MODELS AND OBJECTIVES

Given a $P$-channel signal recorded in a noisy reverberant environment, the physical model in the short-time Fourier transform (STFT) domain can be formulated as

$$\begin{aligned} \boldsymbol{Y}(t,f) &= \boldsymbol{c}(f;p)S_q(t,f) + \boldsymbol{H}(t,f) + \boldsymbol{N}(t,f) \\ &= \boldsymbol{S}(t,f) + \boldsymbol{V}(t,f) \end{aligned} \quad (1)$$

where $S_q \in \mathbb{C}$ is the target speech capture by a reference microphone $q$, $\boldsymbol{c}(f;q) \in \mathbb{C}^{P\times 1}$ is the relative transfer function with the $q^{\text{th}}$ element being one, $\boldsymbol{Y}(t,f)$, $\boldsymbol{c}(f;p)S_q(t,f)$, $\boldsymbol{H}(t,f)$, and $\boldsymbol{N}(t,f) \in \mathbb{C}^{P\times 1}$ respectively represent the STFT vectors of the mixture, direct sound, reverberation, and reverberant noise. We aim at recovering $S_q$ based on $\boldsymbol{Y}$. Our study focuses on dereverberation and the noise is assumed to be an air-conditioning noise, although the proposed approach can be readily applied to deal with more challenging noises. We use $\boldsymbol{S}$ to denote the target speech to extract and $\boldsymbol{V}$ the non-target speech to remove. Our study assumes an offline processing scenario. We normalize the sample variance of each

Following [30]–[32], we train a DNN to directly predict the RI components of the direct-path signal from noisy and reverberant ones via complex spectral mapping. The loss function is

$$\mathcal{L}_{p,\text{RI}} = \left\|\hat{R}_p - \text{Real}(S_p)\right\|_1 + \left\|\hat{I}_p - \text{Imag}(S_p)\right\|_1 \quad (2)$$

where $p$ indexes microphones, $\hat{R}_p$ and $\hat{I}_p$ are the predicted RI components, and $\text{Real}(\cdot)$ and $\text{Imag}(\cdot)$ extract RI components. The enhanced speech is computed as $\hat{S}_p^{(k)} = \hat{R}_p^{(k)} + j\hat{I}_p^{(k)}$. The superscript $k \in \{1,2\}$ denotes it is produced by the $k^{\text{th}}$ DNN (see Figure 1(a)).

Following recent studies [31], [33] that include a magnitude-domain loss for complex spectra approximation, we design the following loss function

$$\mathcal{L}_{p,\text{RI+Mag}} = \mathcal{L}_{p,\text{RI}} + \left\||\hat{S}_p| - |S_p|\right\|_1 \quad (3)$$

The motivation is that using $\mathcal{L}_{p,\text{RI}}$ alone produces worse magnitude estimates, as the estimated magnitudes need to compensate for the estimation error of phase. A major difference from [31], [33] is that we do not perform power or logarithmic compression on the magnitude spectra. This way, the DNN is always trained to estimate an STFT spectrogram that has consistent phase and magnitude structure, and hence would likely produce a good consistent STFT spectrogram at run time [34], [35].

### 3.2. MISO$_1$ System

The multiple-input and single-output system (denoted as MISO$_1$) stacks the RI components of the mixtures at all the microphones and predicts the RI components of the direct-path signal at a reference microphone. This algorithm essentially trains a DNN for non-linear time-varying beamforming. It is simple, fast, and can be easily modified for real-time processing. The model is trained using $\mathcal{L}_{q,\text{RI+Mag}}$.

We emphasize that conventional multi-channel Wiener filtering computes a linear filter per frequency or per T-F unit to project the mixture $Y(t,f)$ onto $S_q(t,f)$, typically based on second-order statistics [1]. In contrast, we utilize a DNN to learn a highly non-linear function to map $Y$ to $S_q$. Although this seems challenging for arrays with arbitrary geometry, for a fixed geometry, this could work as the inter-channel phase patterns are almost fixed for the signal arriving from a specific direction.

### 3.3. MISO$_1$-BF-MISO$_2$ System

The MISO$_1$-BF-MISO$_2$ system includes a MISO network, an MVDR beamformer, and another MISO network. This system is similar to SISO$_1$-BF-SISO$_2$, but we use two MISO networks rather than two SISO networks, since MISO is expected to be better than SISO by doing multi-microphone modeling.

We circularly shift the microphones to estimate the direct-path signal at each microphone. For example, we stack an ordered microphone sequence $<Y_1, ..., Y_P>$ as the inputs to MISO$_1$ to obtain $\hat{S}_1^{(1)}$, and feed in $<Y_p, ..., Y_P, Y_1, ..., Y_{p-1}>$ to obtain $\hat{S}_p^{(1)}$. This strategy would work as we use a circular array with uniformly spaced microphones.

An MVDR beamformer is then computed using $\hat{S}$. The beamforming result $\widehat{BF}_q$ is combined with $Y$ to predict $S_q$ using a MISO network (denoted as MISO$_2$) via complex spectral mapping. This way, post-filtering can also leverage multi-microphone modeling.

### 3.4. MIMO-BF-MISO$_3$ System

The MIMO-BF-MISO$_3$ system consists of a multiple-input and multiple-output (MIMO) network, an MVDR beamformer, and a MISO

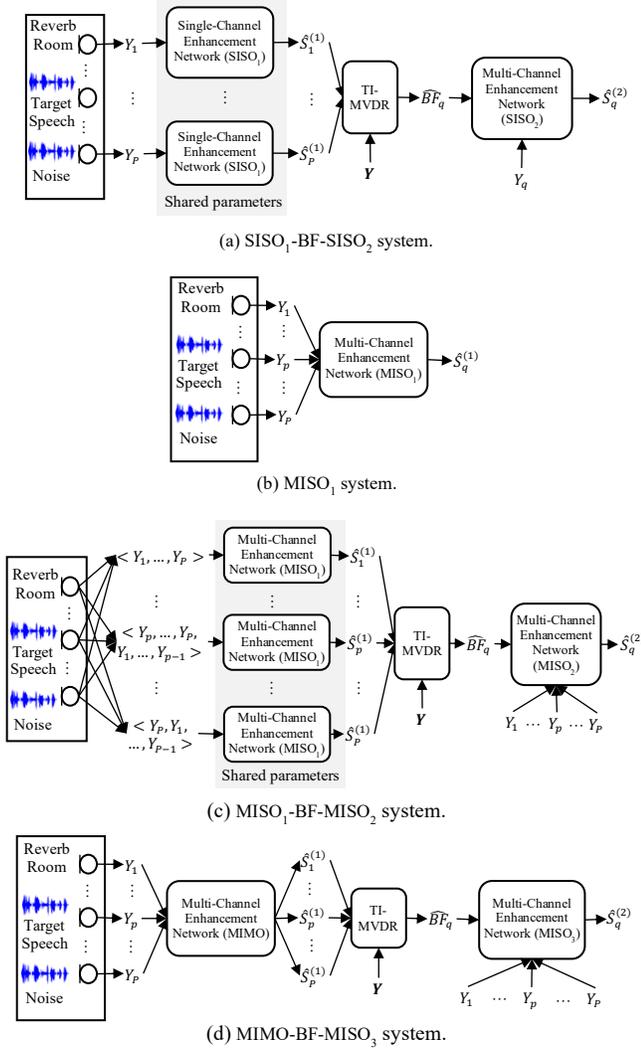

(a) SISO$_1$-BF-SISO$_2$ system.

(b) MISO$_1$ system.

(c) MISO$_1$-BF-MISO$_2$ system.

(d) MIMO-BF-MISO$_3$ system.

Figure 1. System overview.

microphone signal to one before any processing. This can deal with random gains in input. We also assume that the same microphone array is used for training and testing.

## 3. PROPOSED ALGORITHMS

We propose four approaches (denoted as SISO$_1$-BF-SISO$_2$, MISO$_1$, MISO$_1$-BF-MISO$_2$, and MIMO-BF-MISO$_3$, see Figure 1) for multi-channel speech dereverberation. This section discusses each one of them and their combination with beamforming and post-filtering.

### 3.1. SISO$_1$-BF-SISO$_2$ System

The SISO$_1$-BF-SISO$_2$ system contains two single-input and single-output (SISO) networks. The first one (SISO$_1$) performs single-channel complex spectral mapping at each microphone. The enhanced speech is used to compute a time-invariant minimum variance distortion-less response (TI-MVDR) beamformer. The beamforming result $\widehat{BF}_q$ is then combined with the mixture at the reference microphone $Y_q$ as the input to the second SISO network (SISO$_2$) for complex spectral mapping based post-filtering. Note that the trained models of this approach can be applied to arrays with any number of microphones arranged in an unknown geometry.

network. The MIMO network takes in the mixture RI components of all the microphones to predict the RI components of the direct-path signals at all the microphones. This way, we can get an estimate of $S$ for beamforming by performing feed-forwarding only once, rather than $P$ times as in SISO$_1$-BF-SISO$_2$ and MISO$_1$-BF-MISO$_2$. The amount of computation is therefore dramatically reduced. The loss function for the MIMO network is

$$\mathcal{L}_{1,\ldots,P,\text{RI+Mag+PhaseDiff}} = \frac{1}{P}\sum_{p=1}^{P}\mathcal{L}_{p,\text{RI+Mag}} + \frac{1}{P^2-P}\sum_{p'=1}^{P}|S_{p'}|\sum_{p''=1}^{P}\left(1-\cos\left(\angle\hat{S}_{p'}-\angle\hat{S}_{p''}-(\angle S_{p'}-\angle S_{p''})\right)\right)/2 \quad (4)$$

where the second term is a magnitude-weighted cosine distance between the predicted phase differences and the actual phase differences of all the microphone pairs. In our experiments, the second term leads to faster convergence and better performance over using the first term alone.

After obtaining $\hat{S}$, we compute an MVDR beamformer. The beamforming result $\widehat{BF}_q$ is combined with $Y$ to predict $S_q$ using a MISO network (denoted as MISO$_3$) via complex spectral mapping.

### 3.5. MVDR Beamforming

SISO$_1$-BF-SISO$_2$, MISO$_1$-BF-MISO$_2$ and MIMO-BF-MISO$_3$ all have a beamforming module. We use the estimated complex spectra produced by complex spectral mapping to directly compute the speech and noise covariance matrices for TI-MVDR beamforming.

$$\hat{\Phi}^{(s)}(f) = \frac{1}{T}\sum_{t=1}^{T}\hat{S}(t,f)\hat{S}(t,f)^H$$
$$\hat{\Phi}^{(v)}(f) = \frac{1}{T}\sum_{t=1}^{T}\hat{V}(t,f)\hat{V}(t,f)^H \quad (5)$$

where $\hat{V} = Y - \hat{S}$. Previous neural beamforming approaches usually select T-F units dominated by direct sound for covariance matrix estimation by using a T-F mask to compute a weighted sum of all the mixture outer products $Y(t,f)Y(t,f)^H$ within each frequency [8], [10]–[12]. In contrast, we use the estimated complex spectra directly. The rationale is that there may be insufficient T-F units dominated by direct sound especially when room reverberation is very strong, and the phase produced by complex spectral mapping is expected to be better than the mixture phase.

We consider TI-MVDR, as the sound source is assumed to be non-moving within each utterance, and reverberation and the considered noise is largely diffuse. The relative transfer function is computed as follows

$$\hat{r}(f) = \mathcal{P}\{\hat{\Phi}^{(s)}(f)\} \quad (6)$$
$$\hat{c}(f;q) = \hat{r}(f)/\hat{r}_q(f) \quad (7)$$

where $\mathcal{P}\{\cdot\}$ extracts the principal eigenvector [1]. We use Eq. (7) to compute the relative transfer function with respective to a reference microphone $q$.

An MVDR beamformer is computed as

$$\hat{w}(f;q) = \frac{\hat{\Phi}^{(v)}(f)^{-1}\hat{c}(f;q)}{\hat{c}(f;q)^H\hat{\Phi}^{(v)}(f)^{-1}\hat{c}(f;q)} \quad (8)$$

Beamforming results are obtained as $\widehat{BF}_q(t,f) = \hat{w}(f;q)^H Y(t,f)$.

## 4. EXPERIMENTAL SETUP

We use the WSJ0CAM corpus and a large set of simulated room impulse responses (RIRs, in total 39,305 eight-channel RIRs) to simulate room reverberation. See Algorithm 1 for the detailed procedure. For each utterance, we randomly generate a room with different room characteristics, microphone and speaker locations, array configurations, and noise levels. Our study considers an eight-

**Input**: WSJCAM0;
**Output**: spatialized reverberant (and noisy) WSJCAM0;
**For** *dataset, REP* in {*train:5, validation:4, test:3*} set of WSJCAM0 **do**
  **For** each anechoic speech signal $s$ in *dataset* **do**
    **Repeat** *REP* times **do**
    - Draw room length $r_x$ and width $r_y$ from [5,10] m, and height $r_z$ from [3,4] m;
    - Sample mic array height $a_z$ from [1,2] m;
    - Sample array displacement $n_x$ and $n_y$ from [−0.5,0.5] m;
    - Place array center at $\langle\frac{r_x}{2}+n_x,\frac{r_y}{2}+n_y,a_z\rangle$ m;
    - Set array radius $a_r$ to 0.1 m;
    - Sample angle of first mic $\vartheta$ from $[0,\frac{\pi}{4}]$;
    - Place $P(=8)$ mics uniformly on the circle, starting from angle $\vartheta$;
    - Sample target speaker locations: $\langle s_x, s_y, s_z(=a_z)\rangle$ such that distance from target speaker to array center is in between [0.75,2.5] m, and target speaker is at least 0.5 m from each wall;
    - Sample T60 from [0.2,1.3] s;
    - Generate multi-channel impulse responses and convolve them with $s$;
    **If** *dataset* in {*train, validation*} **do**
      -Sample a $P$-channel noise signal $n$ from REVERB training noise;
    **Else**
      - Sample a $P$-channel noise signal $n$ from REVERB testing noise;
    **End**
    - Concatenate channels of reverberated $s$ and $n$ respectively, scale them to an SNR randomly sampled from [5,25] dB, and mix them;
    **End**
  **End**
**End**

Algorithm 1. Data spatialization process.

microphone circular array with the radius fixed at 10 cm. The target speaker is in the same plane as the array, at a distance sampled from [0.75,2.5] m. The training and testing noise (mostly air-conditioning noise) used in the REVERB challenge [36] is utilized to simulate noisy-reverberant mixtures for training and testing, respectively. The reverberation time (T60) is randomly drawn from the range [0.2,1.3] s. The average direct-to-reverberation energy ratio is -3.7 dB with 4.4 dB standard deviation. There are 39,305 (7,861×5, ~80 h), 2,968 (742×4, ~6 h) and 3,264 (1,088×3, ~7 h) eight-channel utterances in the training, validation and test set, respectively.

We validate our algorithms on speech dereverberation using one, two and four microphones. We use the first microphone for the single-microphone task, the first and fifth for the two-microphone task, and the first, third, fifth and seventh for the four-microphone task. Note that the two- and four-microphone setups both have an aperture size of 20 cm. The first microphone is considered as the reference microphone for metric computation. We use scale-invariant SDR (SI-SDR) [37] and PESQ as the evaluation metrics. The former closely reflects the accuracy of estimated magnitude and phase, meaning that estimated magnitude and phase need to compensate with each other to produce a higher SI-SDR, and the latter strongly correlates with the quality of estimated magnitudes.

To evaluate the generalization ability of the trained models, we directly apply them to the recorded data of REVERB [36] for automatic speech recognition (ASR). The recording device is an eight-microphone circular array with 10 cm radius. Note that the array geometry is subject to manufacturing error, which introduces a geometry mismatch between training and testing. The T60 is around 0.7 s and the speaker-to-array distance is 1 m in the near-field case and 2.5 m in the far-field case. We always consider the first microphone as the reference microphone. The ASR backend is built using the most recent Kaldi toolkit.

We use a two-layer BLSTM with convolutional U-Net structure [38], skip connections and dense blocks [39] for dereverberation. See Figure 2 for an illustration of for example the MISO$_2$ network. The rationale for this network design [40] is that BLSTM can model long-range dependencies along time, U-Net can maintain fine-grained structure and exploit large receptive fields, and dense blocks encourage feature re-use and improve the discriminative power of the network. The encoder has one two-dimensional (2D) convolution, and six convolutional blocks, each with 2D convolution, exponential linear units (ELU) and instance normalization (IN), for down-sampling. The decoder contains six convolutional blocks, each with 2D deconvolution, ELU and IN, and one 2D deconvolution, for up-sampling. The RI components of multiple microphones are stacked as feature maps for the network input and output. The window size is 32 ms and hop size 8 ms. The sampling rate is 16 kHz. A 512-point DFT is performed to extract 257-dimensional STFT features at each microphone.

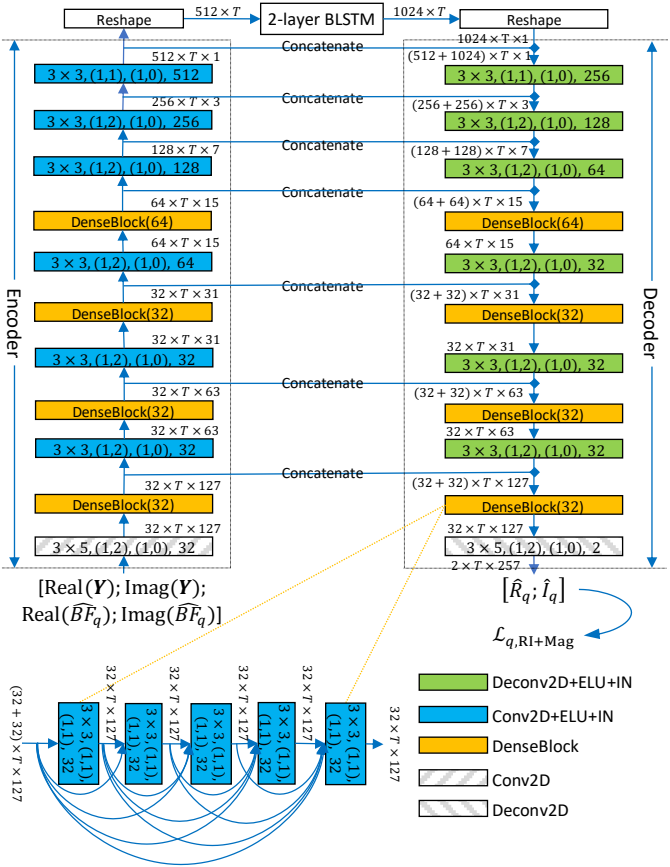

Figure 2. Network architecture of MISO$_2$ for predicting the RI components of $S_q$ from the RI components of $Y$ and $\widehat{BF_q}$. The tensor shape after each block is in the format: *featureMaps×timeSteps×frequencyChannels*. Each Conv2D, Deconv2D, Conv2D+IN+ELU, and Deconv2D+IN+ELU block is specified in the format: *kernelSizeTime × kernelSizeFreq, (stridesTime, stridesFreq), (paddingsTime, paddingsFreq), featureMaps*. Each DenseBlock($g$) contains five Conv2D+IN+ELU blocks with growth rate $g$.

## 5. EVALUATION RESULTS

Table 1 compares the performance of complex spectral mapping with real-valued masking using estimated spectral magnitude mask (SMM) [2] and phase-sensitive mask (PSM) [41] on monaural dereverberation. Much better SI-SDR is obtained using complex spectral mapping based models trained with $\mathcal{L}_{RI}$ and $\mathcal{L}_{RI+Mag}$ over using estimated SMM and PSM, suggesting that complex spectral mapping is effective at phase estimation. In addition, $\mathcal{L}_{RI+Mag}$ leads to much better PESQ than $\mathcal{L}_{RI}$, and slightly better SI-SDR. This indicates the importance of magnitude estimation when PESQ is used as the evaluation metric. The magnitude loss is always included for complex spectral mapping in the following experiments.

Table 1. Average SI-SDR (dB) and PESQ of different methods on monaural dereverberation. $T_a^b(\cdot) = \min(\max(\cdot, a), b)$.

| Methods | SI-SDR | PESQ |
|---|---|---|
| Unprocessed | -3.8 | 1.93 |
| Estimated SMM | 0.6 | 2.92 |
| Estimated PSM | 2.2 | 2.54 |
| $\mathcal{L}_{RI}$ | 6.1 | 2.79 |
| $\mathcal{L}_{RI+Mag}$ | **6.5** | **3.10** |
| Oracle SMM ($T_0^{10}(|S_q|/|Y_q|)$) | 1.5 | 3.39 |
| Oracle PSM ($T_0^1(|S_q|\cos(\angle S_q - \angle Y_q)/|Y_q|)$) | 4.4 | 3.09 |

Table 2. Average SI-SDR (dB) and PESQ of different methods on two- and four-channel dereverberation using simulated test data, and average word error rates (WER) (%) on REVERB real test data.

| Metrics | SI-SDR | | | PESQ | | | WER on REVERB | | |
|---|---|---|---|---|---|---|---|---|---|
| #mics | 1 | 2 | 4 | 1 | 2 | 4 | 1 | 2 | 4 |
| SISO$_1$ | 6.5 | - | - | 3.10 | - | - | 9.62 | - | - |
| SISO$_1$-BF-SISO$_1$ | - | 8.0 | 9.4 | - | 3.20 | 3.29 | - | 8.37 | 7.63 |
| SISO$_1$-BF-SISO$_2$ | - | 8.2 | 10.6 | - | 3.22 | 3.38 | - | 7.96 | 7.25 |
| MISO$_1$ | - | 7.6 | 9.0 | - | 3.22 | 3.33 | - | **7.38** | 6.88 |
| MISO$_1$-BF-MISO$_2$ | - | 8.6 | **10.9** | - | 3.24 | **3.43** | - | **7.38** | **6.30** |
| MIMO | - | 7.2 | 7.8 | - | 3.23 | 3.33 | - | 7.46 | 6.74 |
| MIMO-BF-MISO$_3$ | - | **8.7** | 10.6 | - | **3.28** | 3.41 | - | 7.92 | 6.62 |
| WPE | - | - | - | - | - | - | 14.01 | 13.14 | 11.45 |
| WPE+BeamformIt | - | - | - | - | - | - | - | 12.64 | 9.30 |

Table 2 first reports the enhancement performance of various multi-channel approaches. SISO$_1$ represents a baseline of monaural complex spectral mapping. In SISO$_1$-BF-SISO$_1$, we apply monaural complex spectral mapping on $\widehat{BF_q}$ to estimate target speech $S_q$, while in SISO$_1$-BF-SISO$_2$, complex spectral mapping is applied on the combination of $\widehat{BF_q}$ and $Y_q$ to estimate $S_q$ as in Figure 1(a). SISO$_1$-BF-SISO$_2$ produces better performance than SISO$_1$-BF-SISO$_1$ and SISO$_1$. We emphasize that SISO$_1$-BF-SISO$_1$ represents a typical beamforming followed by post-filtering approach in DNN based multi-channel speech enhancement [4]. In addition, both MISO$_1$ and MIMO are better than SISO$_1$. This indicates that concatenating multiple microphones for complex spectral mapping clearly helps. MIMO is worse than MISO$_1$, because producing multiple outputs is a harder task. Overall, MISO$_1$-BF-MISO$_2$ and MIMO-BF-MISO$_3$ perform the best. This is likely because MISO networks used for post-filtering can benefit from multi-microphone modeling.

In Table 2 we also evaluate the trained models in terms of ASR performance directly on the real test set of REVERB. Both MISO$_1$-BF-MISO$_2$ and MIMO-BF-MISO$_3$ exhibit strong generalization ability, and better ASR performance than SISO$_1$-BF-SISO$_1$ and SISO$_1$-BF-SISO$_2$, which are not sensitive to geometry mismatch. Clear improvements are observed using the trained models over the baseline weighted prediction error (WPE) [36] and WPE followed by BeamformIt algorithms, both available in Kaldi.

## 6. CONCLUSION

We have proposed a multi-microphone complex spectral mapping approach for speech dereverberation, and integrated it with beamforming and post-filtering into a unified system. Experimental results suggest that on a fixed geometry, concatenating multiple microphone signals for complex spectral mapping leads to clear improvements over using a single channel. Future research will consider its extensions to speech enhancement and speaker separation in reverberant and noisy conditions, and investigate its sensitivity to geometry mismatch. We shall also extend the proposed systems to arrays with more than four microphones.